%% file: samplepaper.tex
\documentclass[runningheads]{llncs}
\usepackage[T1]{fontenc}
\usepackage{comment}
\usepackage{graphicx}
\usepackage{subcaption}
\usepackage{xspace}
\usepackage{booktabs}
\usepackage{algorithm}
\usepackage{algpseudocode}
\usepackage{url}
\usepackage{xcolor}

\newcommand{\etal}{\emph{~et~al.\@}\xspace}

\newcommand{\approachname}{\texttt{RAMwavDroid}\@\xspace}

\begin{document}
\title{The Sound of Malware: A Memory Forensics Approach for Android Malware Analysis via Audio Signals}
\titlerunning{The Sound of Malware}
%
%

\author{Silvia Lucia Sanna\inst{1}\orcidID{0009-0002-8269-9777} \and
Massimo Palozzi\inst{2} \and
Leonardo Regano\inst{1}\orcidID{0000-0002-9259-5157} \and
Riccardo Lazzeretti\inst{2}\orcidID{0000-0003-3835-9679} \and
Giorgio Giacinto\inst{1,3}\orcidID{0000-0002-5759-3017}}
\authorrunning{Sanna et al.}
%
\institute{Dip. Ingegneria Elettrica ed Elettronica, \\ Università degli Studi di Cagliari, Cagliari, Italy \and Dip. Ingegneria Informatica, Automatica e Gestionale \\ Università Roma Sapienza, Roma, Italy \and
CINI, Consorzio Interuniversitario Nazionale per l’Informatica, Roma, Italy\\
\email{\{silvial.sanna, leonardo.regano, giogio.giacinto\}}@unica.it \\ \email{\{palozzi.2132174@studenti, riccardo.lazzeretti\}}@uniroma1.it
}

\maketitle              
\begin{abstract}
Android malware analysis is currently facing increasing challenges in achieving robust classification and detecting stealth attacks. Modern threats employ advanced evasion strategies such as code obfuscation, dynamic loading, packing, and even steganographic manipulation of traditional static and dynamic features. These techniques reduce the effectiveness of signature-based systems and degrade the reliability 
of Machine Learning models that depend on explicit semantic indicators such as permissions, API calls, or control-flow structures. In this work, we propose \approachname, a memory forensics malware detection framework that shifts the analysis perspective from semantic program modeling to signal-based structural representation. Both static bytecode and early-execution memory snapshots are transformed into audio waveforms through direct binary-to-waveform mapping, preserving low-level structural patterns without requiring disassembly or feature engineering. The resulting signals are processed using handcrafted spectral descriptors, Convolutional Neural Networks, and transformer-based embeddings. Experiments on CICMalDroid2020 dataset and VirusTotal malware demonstrate that \approachname achieves up to 98.0\% accuracy, outperforming static sonification and competitive state-of-the-art approaches. 

\end{abstract}

\section{Introduction}
\label{sec:intro}
\input{sections/intro}

\section{Background and Related Works}
In this section, we present the main concepts of Android Malware analysis, focusing on traditional and memory forensics techniques. Lastly, we present current literature on the binary-to-audio encoding.
\label{sec:backsota}
\input{sections/background}


\section{Methodology}
\label{sec:methodology}

\begin{figure}
    \centering
    \includegraphics[width=\linewidth]{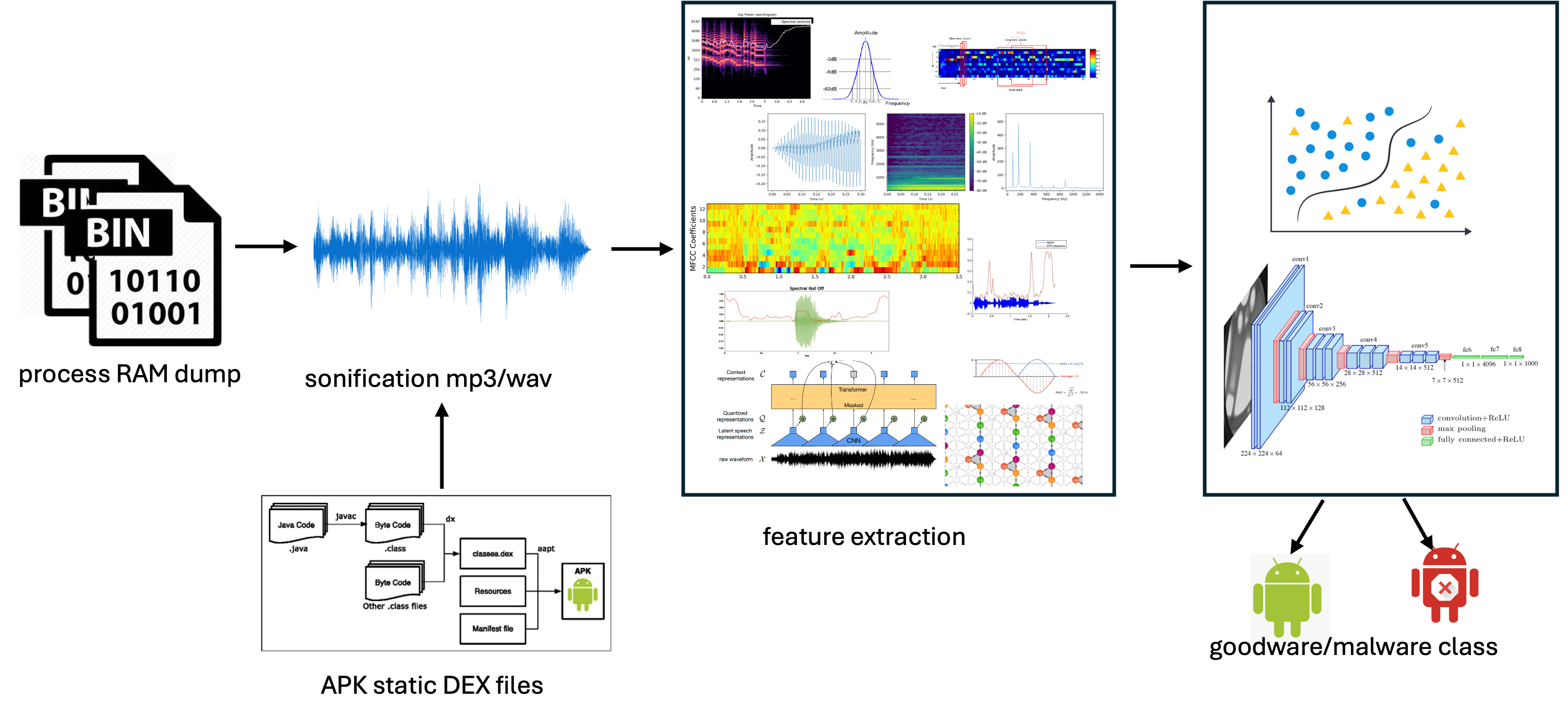}
    \caption{\approachname analysis pipeline comparing static DEX files and early-execution memory dumps through audio encoding. Both representations are sonified (WAV 8-bit, WAV 16-bit, and MP3) and processed via ML models and CNN for classification.}
    \label{fig:placeholder}
\end{figure}

In this section, we present \approachname, illustrated in Figure \ref{fig:placeholder}, our methodology to detect Android malware from static and dynamic analysis via audio signal encoding.

\input{sections/methodology}

\section{Results}
\label{sec:results}
In the following, we present the main results of our protocol by first comparing \approachname with other state-of-the-art tools based on the sonification, memory forensics and malware detection on the CICMalDroid2020 dataset~\cite{maldroid2020dataset,cicmaldroid} according to static and dynamic analysis. Here we highlight the different sonification algorithms (i.e., \texttt{WAV8}, \texttt{WAV16}, \texttt{MP3}) and classification methodologies (i.e., feature extraction, CNNs, \texttt{Wav2Vec2}). Subsequently, we present the results on a stealthy malware dataset downloaded from VirusTotal, stressing the power of memory forensics and the benefit of audio sonification encoding.
\input{sections/results}


\section{Conclusions}
\label{sec:conclusion}
This work presented \approachname, a memory-forensics-driven Android malware detection framework based on audio signal encoding. By transforming both static bytecode and early-execution memory snapshots into waveform representations, we demonstrated that structural properties of binary data can be effectively analyzed using signal processing and ML techniques.

Experimental results show that static sonification achieves competitive performance, while dynamic memory sonification significantly improves separability, reaching 98.0\% accuracy on CICMalDroid2020. The combination of application code and execution-context regions proved more informative than isolated memory components, highlighting the importance of runtime artifacts. Furthermore, the comparison of encoding formats revealed that discriminative information is preserved under lossy compression, suggesting that coarse-grained structural patterns dominate classification performance. The convergence of results across traditional ML models, CNNs, and transformer-based embeddings indicates that sonified runtime memory contains stable and learnable patterns. These findings position dynamic memory sonification as a promising complementary approach to traditional static and dynamic analysis techniques.

Future work will focus on scaling the evaluation to larger datasets, investigating robustness against advanced obfuscation strategies, and exploring multi-temporal memory acquisition to further capture behavioral evolution during execution.

\begin{credits}
\subsubsection{\ackname} Removed for anonymization.

\subsubsection{\discintname}
The authors have no competing interests. 

\subsubsection{Code Data Availability}
The code is available at: \url{https://anonymous.4open.science/r/RAMwavDroid-7401} while the dataset is available from the corresponding author.
\end{credits}

\bibliographystyle{splncs04}
\bibliography{biblio}

\end{document}

%% file: sections/intro.tex
The widespread adoption of Android devices has been accompanied by a parallel increase in Android-targeted malware~\cite{Alrammal22_JIT}. Traditional detection mechanisms have evolved from signature-based approaches to Machine Learning (ML) models leveraging static and dynamic features, respectively without and with the execution of the target app~\cite{Bensaoud2024,Islam_ITCPS23,Ruggia_ACSAC21,Ruggia_ACM24}. Static analysis techniques, such as permission inspection and API-call modeling, are computationally efficient but can be evaded through code obfuscation, packing, and dynamic loading~\cite{Soi_JISA24,drebin,Fereidooni2016}. Dynamic analysis techniques, including runtime monitoring and information-flow tracking, improve resilience but suffer from execution overhead, incomplete behavioral coverage, and susceptibility to anti-analysis techniques~\cite{mahindru2017,asghari2025unpackunpacklivingpackers,Keyes21_RDAAPS,Valerian_EuroS&PW24}. Memory forensics offers a complementary perspective by capturing volatile runtime artifacts that are not visible in packaged application binaries~\cite{Khalid_volmemdroid_ESA24,sanna2026explainablememoryforensicsapproach,Naeem23_ESA,Bellizzi_Access22}. Process-level memory snapshots may include decrypted payloads, dynamically loaded libraries, runtime-generated code, and execution-context data structures. These artifacts provide valuable insight into malware behavior, particularly during early execution stages~\cite{kumar2023inviseal}. In parallel, recent research has explored alternative binary representations for malware detection, including image-based and audio-based encodings~\cite{Daoudi21_Springer,didroid,Aligombe_DFRWS23_crgbmem,farrokhmanesh2016,Walden2022,mercaldo2021}. Audio-based approaches transform binary data into waveform signals, enabling the use of signal processing and Deep Learning (DL) models originally developed for speech and music analysis. However, existing audio-based Android malware detection methods primarily focus on static binaries~\cite{kural2023,Casolare2021,tarwireyi2023}.

In this work, we bridge memory forensics and audio encoding by proposing \approachname, a unified sonification framework that supports both static APK bytecode and dynamic runtime memory snapshots. Conceptually, sonification performs a modality shift from program analysis to signal analysis. While it does not replace traditional static or dynamic analysis, it provides a complementary, signal-based perspective on malware structure~\cite{tarwireyi2023}. Our representation is built directly from raw byte sequences and runtime memory contents, thus avoiding dependence on fragile handcrafted semantic abstractions and deobfuscation steps. By directly mapping raw bytes to waveform amplitudes, \approachname preserves low-level structural patterns while enabling spectral analysis and deep representation learning. The approach is evaluated using multiple audio encodings, feature extraction paradigms, classification models and datasets.

Our contributions are threefold. First, we introduce a dynamic memory sonification pipeline that integrates process-level memory acquisition with binary-to-audio encoding. Secondly, we systematically compare encoding formats and memory-region selection strategies to identify discriminative runtime artifacts. Thirdly, we demonstrate that dynamic sonification achieves up to 98.0\% accuracy surpassing static sonification and competitive state-of-the-art approaches. These results indicate that structural patterns embedded in runtime memory provide a robust signal representation for Android malware detection.

The remainder of this paper is structured as follows. Section~\ref{sec:backsota} introduces the main concepts of Android APK and malware analysis, memory forensics and audio encoding. Our proposed methodology is presented in Section~\ref{sec:methodology} with the corresponding results in Section~\ref{sec:results}, while Section~\ref{sec:conclusion} closes the paper.

%% file: sections/background.tex
\subsection{Android APK}
Android applications are distributed as Android Packages (APKs), which are structured ZIP archives containing executable bytecode, resources, native libraries, and metadata. The core component of an APK is the \texttt{classes.dex} file, which stores Dalvik bytecode executed by the Android Runtime (ART). ART supports both just-in-time (JIT) and ahead-of-time (AOT) compilation, garbage collection, and runtime profiling, directly influencing the layout of memory during execution. The \texttt{AndroidManifest.xml} defines application components and permissions, while the \texttt{META-INF/} directory contains cryptographic signatures that ensure application integrity. At runtime, Android applications execute within isolated Linux processes. The virtual memory mappings of each process are stated in \texttt{/proc/<pid>/maps}, which enumerates heap, stack, shared libraries, JIT-compiled regions, and other memory segments. 

\subsection{Android Malware}

Android malware detection has progressively evolved from signature-based techniques to advanced ML systems. Drebin~\cite{drebin} demonstrated that lightweight static features such as permissions and API calls can effectively distinguish malicious applications using linear classifiers, while preserving explainability. Hybrid approaches such as SAMADroid~\cite{Arshad18_Access} combined static and dynamic features into multi-level detection frameworks. The survey by Faruki\etal~\cite{Faruki23_Inf} analyze the Android threat landscape, highlighting the increasing use of obfuscation, dynamic code loading, and native libraries to evade detection. More recent works focus on explainability and structural modeling. For example, Soi\etal~\cite{Soi_JISA24} proposed a malware detection approach based on function call graphs to enhance interpretability. Dynamic analysis techniques monitor runtime behavior rather than static artifacts. TaintDroid~\cite{Enck14_TaintDroid,wong_intellidroid_2016} introduced real-time information flow tracking on Android to detect privacy leaks. UI-driven dynamic behavior analysis has also been explored to trigger hidden malicious behaviors~\cite{zheng2012_UIBased}. Although dynamic methods improve robustness against obfuscation, they introduce performance overhead and partial code coverage limitations.

\subsection{Android Memory Forensics}

Memory forensics enables the analysis of volatile runtime artifacts that cannot be observed through static inspection. Full physical memory acquisition on Android devices is commonly performed using LiME (Linux Memory Extractor)~\cite{504ensicsLabsLiME}, an open-source kernel module that captures complete RAM images, including kernel and user-space artifacts. LiME integrates with forensic frameworks such as \texttt{Volatility}, offering reproducible and system-wide visibility. However, full memory acquisition requires root privileges and kernel module insertion. 

To reduce intrusiveness, user-space approaches such as Fridump~\cite{Nightbringer21Fridump} leverage the Frida instrumentation framework to extract process-level memory regions without modifying the kernel. This methodology allows selective acquisition of readable memory mappings, preserving runtime artifacts such as decrypted payloads and dynamically loaded code. VolMemDroid~\cite{Khalid_volmemdroid_ESA24} further systematizes Android volatile memory acquisition by providing a structured and reproducible methodology for collecting and analyzing RAM artifacts across different Android versions and device configurations, improving consistency and forensic soundness.

Recent works integrate memory forensics with ML. DroidScraper~\cite{AliGombe_RAID19_droidscraper} demonstrated in-memory object recovery and reconstruction for Android applications. The work crgb\_mem~\cite{Aligombe_DFRWS23_crgbmem} explored the intersection between memory forensics and machine learning, showing that volatile memory representations can be processed using data-driven techniques. Vedrando~\cite{Bellizzi_JCP23_vedrando} revealed stealthy Android attack steps through memory-forensic analysis. More recently, AI-driven approaches have been proposed to enhance robustness and interpretability in digital forensic workflows~\cite{sanna2026explainablememoryforensicsapproach}. These works highlight that memory snapshots provide access to transient, decrypted, and dynamically generated artifacts that are invisible in static APK analysis, offering a richer substrate for malware detection and forensic investigation.

\subsection{Audio Encoding for Malware Detection}

Signal-processing-based malware detection transforms binary data into alternative representations such as images or audio. Early audio-based malware detection was proposed by Farrokhmanesh and Hamzeh~\cite{farrokhmanesh2016}, who converted binaries into audio signals and extracted Mel-Frequency Cepstral Coefficients (MFCCs) features for classification. Within the Android ecosystem, audio-based detection approaches convert DEX or APK files into waveform representations and extract spectral features such as MFCCs, chroma, spectral centroid, spectral bandwidth, zero-crossing rate, and RMS energy. Multi-feature fusion techniques combined with ensemble models have demonstrated high classification performance on benchmark datasets. DL approaches further extend this paradigm. CNNs effectively process spectrogram representations, while self-supervised learning models such as \texttt{Wav2Vec2}~\cite{Baevski2020} learn contextual embeddings directly from raw waveforms without requiring manual feature engineering. Transfer learning improves generalization when malware datasets are limited. Despite their strong performance, most audio-based approaches rely on static binaries. Integrating memory forensics with audio encoding enables dynamic sonification of runtime memory snapshots, capturing decrypted code, runtime-loaded libraries, and volatile artifacts. This integration bridges static structural representations and behavioral runtime analysis, potentially improving robustness against obfuscation and stealth techniques.

%% file: sections/methodology.tex
\subsection{Dataset Creation}
\approachname supports static and dynamic analysis within a unified processing pipeline. In the static configuration, each APK produces a single audio representation derived from its bytecode. In the dynamic configuration, multiple audio signals may be generated from different memory regions of the same application. These signals are subsequently aggregated at the feature level to form a unified representation of the application. This dual setup enables a direct comparison between pre-execution artifacts and runtime information. While static sonification captures structural properties of packaged code, dynamic sonification incorporates transient and execution-dependent artifacts that may not be observable through static inspection alone. The unified pipeline ensures that differences in performance can be attributed to the information source rather than to methodological inconsistencies.

\textbf{Static Dataset}. The static dataset consists of APKs collected from publicly available repositories and labeled as benign or malicious. Each APK is treated as a compressed archive from which the Dalvik bytecode (\texttt{classes.dex}) is extracted. This file contains the application executable and serves as the input for static sonification. By operating directly on bytecode, the static dataset enables evaluation of malware detection performance using only pre-execution artifacts. 

\textbf{Dynamic Dataset}. To incorporate runtime behavior, we constructed a dynamic dataset composed of process-level memory snapshots acquired using \texttt{Fridump}~\cite{Nightbringer21Fridump}, a user-space memory extraction framework built on top of the \texttt{Frida} dynamic instrumentation engine~\cite{FridaAndroid}. \texttt{Fridump} enables selective acquisition of a running application’s virtual address space without requiring kernel modification or full physical memory imaging. By operating entirely in user space, it attaches to the target process and leverages \texttt{Frida}’s introspection APIs to enumerate and read memory regions in a controlled manner. Each APK was executed in a standardized environment with a fixed system configuration. 
Applications were launched using an automated procedure that ensured consistent initialization 
(e.g., installation, permission granting, and immediate startup), and memory acquisition 
was triggered at a fixed point shortly after process initialization. 
Experiments were conducted using the \texttt{Genymotion} Android emulator~\cite{genymotion}, which provides a stable 
virtualized environment while exhibiting fewer artifacts typically associated with analysis 
sandboxes (e.g., root indicators or debugging traces) compared to the standard Android Studio 
emulators. This choice reduces the likelihood of triggering anti-debugging or anti-analysis 
mechanisms implemented by certain malware samples, thereby enabling more faithful runtime observation.

\textbf{Memory Acquisition}. RAM extraction was triggered immediately after the APK was loaded and the application process completed its initialization phase. At this stage, core components such as Dalvik/ART runtime structures, application bytecode, native libraries, and essential heap and stack regions are already mapped into memory. Capturing the snapshot at this early execution point allows us to include runtime-dependent artifacts while minimizing variability caused by prolonged interaction or background activity. \texttt{Fridump} first parses the \texttt{/proc/<PID>/maps} interface to obtain a complete description of the process memory layout, including virtual address ranges, access permissions, and associated file mappings. Based on this information, only regions marked as readable are extracted. Each region is acquired in bounded chunks to ensure stability and avoid disrupting the running process. The acquisition procedure produces two complementary artifacts: \emph{(i)} a memory map describing the full virtual address space of the process, and \emph{(ii)} a set of raw binary dumps corresponding to each readable mapping. Unlike full physical memory acquisition tools, this process-scoped strategy focuses exclusively on the address space of the target application, thereby reducing noise from unrelated system components while preserving relevant runtime artifacts. These include loaded bytecode segments, shared libraries, stack frames, heap allocations, and potentially decrypted or dynamically generated content that would not be observable through static APK analysis. To facilitate systematic evaluation, we defined multiple region-selection strategies targeting specific runtime components, such as execution stack segments, application binaries (e.g., \texttt{base.apk}, \texttt{.odex}, \texttt{.vdex}), shared libraries, Dalvik runtime regions, and application data segments. This modular design enables controlled experimentation on the discriminative contribution of different memory areas while maintaining a reproducible and minimally intrusive acquisition workflow.

\textbf{Sonification and Encodings}. Sonification constitutes the central methodological contribution of \approachname, enabling a shift from traditional program analysis to signal-based representation.
Instead of modeling explicit semantic constructs such as API calls, permissions, or control-flow graphs, \approachname leverages distributional and structural properties embedded in raw data. It is agnostic to programming language and compilation strategy, does not require deobfuscation, and enables the direct application of established audio-analysis and deep-learning techniques.
Rather than relying on syntactic parsing, opcode inspection, or behavioral tracing, we reinterpret binary data as a structured numerical signal and analyze it through the framework of audio processing. In this perspective, both executable code and runtime memory are treated as ordered byte sequences whose statistical and structural regularities can be captured in the temporal and spectral domains, mapping continuous data. The transformation is performed through direct binary-to-waveform mapping. Given a sequence of bytes $b_i \in [0,255]$, each value is interpreted as an audio sample amplitude, generating a mono-channel waveform in which the temporal progression directly reflects the sequential layout of the binary stream, as detailed in Algorithm~\ref{alg:sonification}. Importantly, no disassembly, normalization, semantic filtering, or opcode-level abstraction is applied before conversion. This design choice intentionally preserves low-level structural properties embedded in the binary data, including repetition patterns, alignment artifacts, padding regions, packed segments, encrypted payloads, and memory layout characteristics. At the same time, it enables the application of mature signal-processing techniques.

\input{sections/algorithm_sonification}

Unlike traditional static analysis, which extracts handcrafted syntactic or semantic program features, sonification treats malware as a signal. Executable files are not random byte collections; they exhibit non-uniform statistical distributions induced by instruction sets, control-flow patterns, constant pools, compiler artifacts, and runtime structures. When mapped to an audio waveform, these regularities manifest as measurable spectral signatures. Repeated byte sequences may produce frequency concentration bands, aligned instruction blocks may induce harmonic-like structures, while encrypted or compressed regions often generate broadband spectral behavior. Temporal discontinuities can reflect transitions between memory regions or structural boundaries in the binary layout. In the static setting, the input corresponds to the Dalvik bytecode extracted from the APK, producing a signal that encodes the packaged application structure. In the dynamic setting, the input consists of selected runtime memory regions acquired during application startup. Memory snapshots contain execution-context information such as decrypted payloads, dynamically loaded libraries, just-in-time compiled code, heap allocations, and stack artifacts. By sonifying these regions, the waveform becomes a compact signal-level abstraction of the application’s operational state. This extension allows the capture of transient and execution-dependent artifacts that may evade purely static inspection, while preserving a uniform processing pipeline across static and dynamic scenarios.

To assess the robustness and fidelity of this representation, we experiment with multiple audio encodings. Lossless formats (e.g., WAV) preserve an exact byte-to-sample correspondence, ensuring maximal structural fidelity and maintaining a one-to-one mapping between the binary stream and waveform amplitude values. In contrast, lossy encodings (e.g., MP3) apply perceptual compression mechanisms that selectively remove components considered irrelevant to human auditory perception. From a cybersecurity standpoint, this compression acts as a structured perturbation of the signal. Evaluating classification performance under lossy encoding allows us to determine whether malware-discriminative information is encoded in stable structural patterns or in fine-grained byte-level variations. The comparison between encodings therefore serves a dual purpose: \emph{(i)} assessing the stability of malware signatures under signal degradation, and \emph{(ii)} analyzing the trade-off between representational precision, storage efficiency, and computational cost. 
The resulting audio representations form the foundation for subsequent feature extraction and classification stages, enabling the systematic evaluation of both handcrafted audio descriptors and learned deep representations.

\subsection{Classification}
In this section, we present the methodology used to classify the audio signals.

\textbf{Feature Extraction}. After sonification, audio signals are transformed into numerical representations suitable for ML algorithms. We adopt two complementary lines for feature extraction. \textit{The first strategy} relies on established audio descriptors commonly used in signal processing and music information retrieval, for example in music emotion recognition~\cite{yang2012mfcc}. These handcrafted features summarize spectral, temporal, and cepstral properties of the waveform. Examples include frequency-domain statistics, energy-related descriptors, and perceptually motivated coefficients such as MFCCs. Together, these features capture distributional properties of the signal, frequency concentration patterns, harmonic content, and temporal variability. This approach provides an interpretable and compact representation that can be directly fed into conventional classifiers. The \textit{second approach} leverages learned representations extracted from a pre-trained transformer model. Instead of manually defining signal descriptors, the raw waveform is processed by a DNN trained through self-supervised learning. The resulting embeddings encode high-level temporal and spectral dependencies learned from large-scale audio corpora. These embeddings serve as dense feature vectors that capture complex signal structure without requiring manual feature engineering. By combining handcrafted descriptors and learned embeddings, we evaluate both classical signal-processing features and modern representation learning approaches within the same framework.

\textbf{Convolutional Neural Networks}. In addition to traditional classifiers, we evaluate a CNN tailored for audio-based classification. The CNN operates on structured audio representations, such as time–frequency matrices or feature maps, and learns hierarchical patterns directly from the data. Convolutional layers extract local patterns along temporal and frequency dimensions, enabling the model to detect recurring spectral patterns, abrupt transitions, and structural regularities within the sonified malware signal. Deeper layers progressively aggregate these local patterns into higher-level abstractions that represent discriminative characteristics of benign and malicious applications. Unlike fixed handcrafted descriptors, CNNs learn task-specific filters optimized for classification. This property makes them particularly suitable for capturing subtle structural differences between malware and benign samples. The CNN architecture, therefore, serves as a baseline capable of modeling nonlinear relationships that may not be captured by traditional ML algorithms.

\textbf{Wav2Vec}. To further explore representation learning, we employ a pre-trained \texttt{Wav2Vec2} model as an audio feature extractor~\cite{Baevski2020}. \texttt{Wav2Vec2} is trained using self-supervised objectives on large-scale unlabeled speech data and learns contextual representations directly from raw waveforms. The model combines convolutional encoders with transformer layers, enabling it to capture both local signal patterns and long-range temporal dependencies. Within \approachname, \texttt{Wav2Vec2} is used in a transfer-learning setting. The sonified malware waveform is fed into the model, and high-dimensional contextual embeddings are extracted from the final hidden layer. These embeddings are then used as input to lightweight downstream classifiers. This strategy leverages knowledge acquired from large general-purpose audio corpora to encode structural properties of malware signals. By avoiding training from scratch, \texttt{Wav2Vec2} enables efficient learning even with limited labeled malware samples. The comparison between CNN-based learning and transformer-based embeddings allows us to assess whether self-supervised audio representations can generalize effectively to cybersecurity-oriented signal classification tasks.

%% file: sections/algorithm_sonification.tex
\begin{algorithm}[p]
\caption{Binary-to-Audio Sonification (Static \& Dynamic)}
\begin{algorithmic}[1]
\Require Dataset type $\mathcal{T}\in\{\mathrm{Static},\mathrm{Dynamic}\}$;
        input source $\mathcal{S}$;
        encoding $\in\{\mathrm{WAV8},\mathrm{WAV16},\mathrm{MP3}\}$;
        sample rate $f_s$
\Ensure Audio file(s) $\mathcal{A}$

\Statex
\Comment{Byte Extraction}
\If{$\mathcal{T} = \mathrm{Static}$}
    \State Extract \texttt{classes.dex} from APK archive
    \State $\mathcal{B} \gets \{b_1,\dots,b_n\}$ from dex file
\ElsIf{$\mathcal{T} = \mathrm{Dynamic}$}
    \ForAll{selected memory regions $r$ in $\mathcal{S}$}
        \State Read raw dump file $r$\texttt{\_dump.data}
        \State $\mathcal{B}_r \gets \{b_1,\dots,b_{n_r}\}$
    \EndFor
\EndIf

\Statex
\Comment{Direct Binary-to-Waveform Mapping}
\ForAll{byte sequence $\mathcal{B}$}
    \For{$i \gets 1$ to $n$}
        \State $s[i] \gets b_i$ 
        \Comment{Amplitude sample $s_i \in [0,255]$}
    \EndFor
\EndFor

\Statex
\Comment{Audio Container Construction}
\If{encoding = WAV8}
    \State channels $\gets 1$
    \State sample\_width $\gets 1$ byte
    \State Write $s[i]$ directly as 8-bit PCM frames
\ElsIf{encoding = WAV16}
    \State channels $\gets 1$
    \State sample\_width $\gets 2$ bytes
    \State Write $s[i]$ as 16-bit PCM frames
\ElsIf{encoding = MP3}
    \State Convert $s[i]$ to mono audio segment
    \State Apply perceptual compression (MP3 encoder)
\EndIf

\Statex
\Comment{Output Organization}
\If{$\mathcal{T} = \mathrm{Static}$}
    \State Save one audio file per APK
\ElsIf{$\mathcal{T} = \mathrm{Dynamic}$}
    \State Save one audio file per memory region
\EndIf

\State \Return $\mathcal{A}$
\end{algorithmic}
\label{alg:sonification}
\end{algorithm}

%% file: sections/results.tex
\subsection{Static Analysis}

We first evaluated the proposed sonification pipeline on the CICMalDroid2020 dataset using static \texttt{classes.dex} extraction. This stage served to validate the correctness of the implementation before extending the methodology to dynamic memory analysis. The best-performing static configuration was obtained using \texttt{WAV8} encoding combined with manual audio feature extraction through \texttt{librosa} and Random Forest classification, achieving $93.9$\% overall accuracy. The SMS category achieved the highest F1-score ($98.8$\%), while Adware and Banking samples were more challenging, with F1-scores of $82.4$\% and $85.5$\%, respectively. Increasing the bit depth to \texttt{WAV16} did not improve classification performance ($92.3$\% accuracy). Similarly, \texttt{MP3} encoding achieved $92.6$\% accuracy while reducing storage footprint (approximately $25$\%) and processing time.

When replacing manual descriptors with \texttt{Wav2Vec2} embeddings using \texttt{WAV8} encoding, overall accuracy decreased slightly to $91.5$\%\footnote{We did not experiment with \texttt{Wav2Vec2} embeddings for the \texttt{WAV16} and \texttt{MP3} encodings, since on the one hand we noted an accuracy decrease for all except one malware category when adopting the \texttt{Wav2Vec2} embeddings with the \texttt{WAV8} encoding, and on the other hand we obtained an accuracy decrease for all malware categories when adopting the \texttt{WAV16} and \texttt{MP3} encodings with manual audio feature extraction with respect to \texttt{WAV8} encoding.}. However, feature extraction time was significantly reduced (approximately $2$ hours compared to $6-7$ hours for manual features), demonstrating the computational advantage of transformer-based representations. 

\begin{figure}
    \centering
    \includegraphics[width=\linewidth]{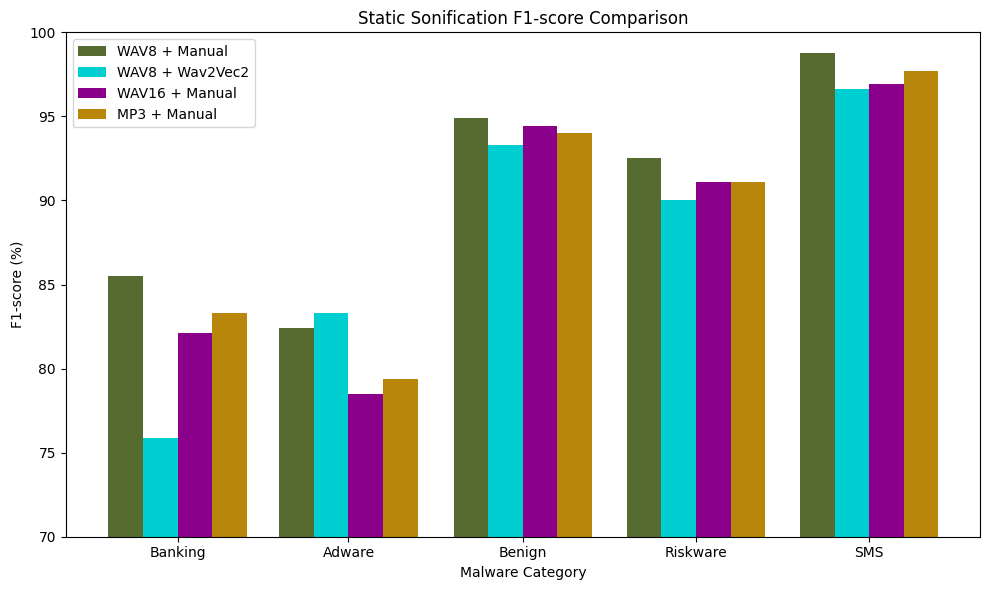}
    \caption{Static Sonification Results}
    \label{fig:staticsonif}
\end{figure}

These results, detailed in Figure~\ref{fig:staticsonif}, suggest that higher amplitude resolution does not introduce additional discriminative information and that moderate lossy compression preserves relevant structural characteristics of the binary data. Overall, static sonification produced results comparable to prior audio-based approaches and confirmed the validity of the implemented system before proceeding to dynamic experiments.

\subsection{Dynamic Analysis}

Dynamic memory sonification represents the central contribution of this work. 
Experiments were conducted on 600 Android applications ($300$ benign and $300$ malicious) from CICMalDroid, using process-scoped memory dumps acquired immediately after application initialization, as described in the methodology. Each snapshot captured the address space of the target process at early execution, including loaded bytecode segments, shared libraries, stack frames, heap allocations, and runtime-generated structures. The classification is performed at APK level by aggregating the memory-region results. In order to evaluate the impact of audio representation on runtime artifacts, we combined three audio encodings, ten memory-region selection strategies, and ten classification models. The primary objective of this phase was to determine which encoding format provides the most effective representation for dynamic memory sonification, while simultaneously analyzing the contribution of different runtime regions.

The highest performance achieved was $98.0$\% accuracy, with zero false positives and one false negative. This performance (Table~\ref{tab:memory_region_accuracy}) was reached by seven region-selection strategies, including \texttt{apk\_full\_stack}, \texttt{apk\_full}, \texttt{apk\_full\_memory}, \texttt{base\_apk\_stack}, \texttt{base\_apk\_memory}, \texttt{data\_stack}, and \texttt{all\_data\_paths}.  These strategies share a common characteristic: we combine application code segments (e.g., \texttt{base.apk}, \texttt{.odex}, \texttt{.vdex}) with execution context components such as stack frames or runtime-loaded data. The consistent peak performance across these configurations suggests that discriminative information is not confined to a single memory component, but emerges from the interaction between static code and runtime execution structures. In contrast, isolated memory sections yielded lower accuracy. The \texttt{stack}-only strategy achieved $90.0$\%, indicating that execution context alone does not provide sufficient structural information for reliable classification. Similarly, the \texttt{complete\_memory\_regions} strategy reached $93.0$\%. Although this configuration aggregates all available mappings, its lower performance suggests that indiscriminately including heap and auxiliary regions introduces noise rather than additional discriminative content. The \texttt{base\_apk} selection alone achieved $97.0$\%, confirming that static code remains highly informative, yet slightly less effective than configurations that also incorporate runtime context. Among all strategies, \texttt{apk\_full\_stack} proved particularly stable across classifiers. Multiple models, including Random Forest, boosting methods, CNN1D, and ensemble approaches, converged to identical performance. This stability indicates that once informative runtime sections are selected, classification performance becomes largely model-independent.

\begin{table}[t]
\caption{Best classification results for each memory region selection strategy.}
\label{tab:memory_region_accuracy}
\centering
\begin{tabular}{lllccc}
\toprule
Strategy & Encoding & Model & Accuracy (\%) & FP & FN \\
\midrule
all\_data\_paths & MP3 & Random Forest & 98.0 & 0 & 1 \\
apk\_full & MP3 & Random Forest & 98.0 & 0 & 1 \\
apk\_full\_memory & MP3 & Random Forest & 98.0 & 0 & 1 \\
apk\_full\_stack & MP3 & Random Forest & 98.0 & 0 & 1 \\
base\_apk\_stack & MP3 & Random Forest & 98.0 & 0 & 1 \\
base\_apk\_memory & MP3 & Random Forest & 98.0 & 0 & 1 \\
data\_stack & MP3 & Random Forest & 98.0 & 0 & 1 \\
base\_apk & WAV16 & XGBoost & 97.0 & 0 & 2 \\
complete\_memory\_regions & WAV16 & CatBoost & 93.0 & 1 & 3 \\
stack & WAV8 & Random Forest & 90.0 & 5 & 1 \\
\bottomrule
\end{tabular}
\end{table}

When analyzing average performance across all configurations, \texttt{WAV8} achieved the highest mean accuracy ($95.8$\%), followed by \texttt{WAV16} ($93.3$\%) and \texttt{MP3} ($92.6$\%), see Table~\ref{tab:encoding_comparison}. However, all three encodings reached $98.0$\% accuracy under optimal region selection. This result is particularly significant: although \texttt{WAV8} preserves a direct byte-to-sample mapping, lossy \texttt{MP3} compression does not eliminate the structural regularities that differentiate benign from malicious memory. 

\begin{table}[t]
\caption{Average classification performance for each audio encoding format across all configurations.}
\label{tab:encoding_comparison}
\centering
\small
\begin{tabular}{lcccc}
\toprule
Encoding & Avg Accuracy (\%) & Avg F1-Score (\%) & Avg FN & Avg FP \\
\midrule
WAV8 & 95.8 & 95.8 & 1.64 & 0.81 \\
WAV16 & 93.3 & 93.3 & 2.56 & 1.39 \\
MP3 & 92.6 & 92.4 & 3.24 & 1.07 \\
\bottomrule
\end{tabular}
\end{table}

These findings suggest that dynamic memory sonification primarily relies on coarse-grained structural patterns embedded in runtime artifacts rather than fine-grained amplitude precision. Consequently, discriminative information is robust under moderate compression and does not depend on high bit-depth representation. Overall, dynamic analysis demonstrates that runtime memory provides additional separability compared to static APK sonification. The combination of execution context and application code captures behavioral artifacts that are not observable in packaged bytecode alone, thereby improving classification robustness while maintaining representation efficiency.

\subsection{Comparison with the state of the art}

\begin{figure}
    \centering
    \includegraphics[width=\linewidth]{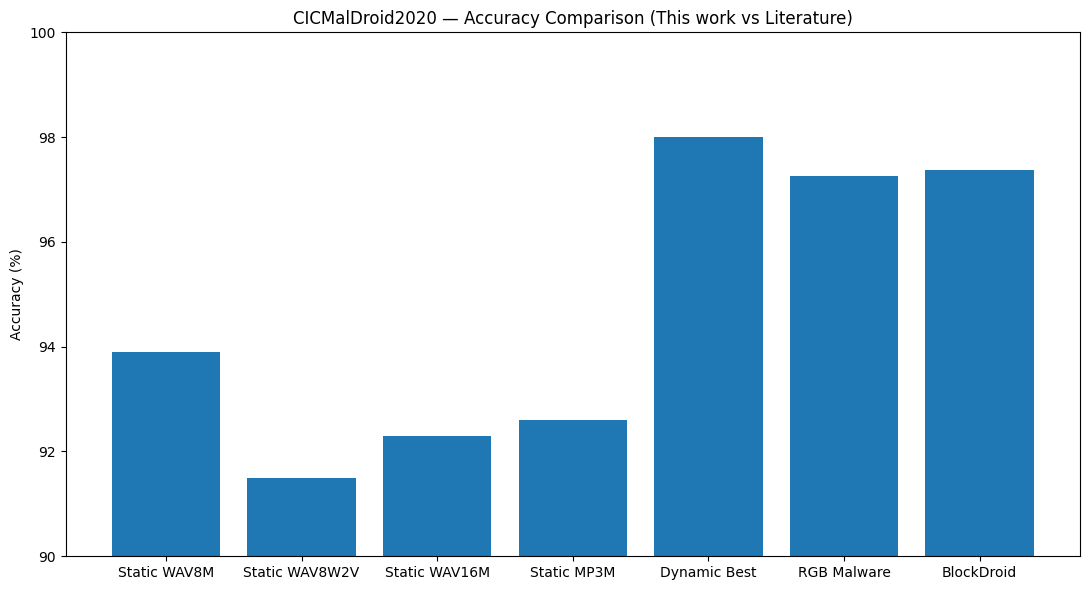}
    \caption{State-of-the-art-comparison}
    \label{fig:literature}
\end{figure}

Figure~\ref{fig:literature} compares \approachname with representative state-of-the-art methods evaluated on the CICMalDroid2020 dataset, namely an image-based approach~\cite{wang2024androidmalwaredetectionbased} and BlockDroid~\cite{BlockDroid2024}, a CNN-based detector. Static sonification achieves competitive performance across different encodings, with \texttt{WAV8} combined with manual descriptors reaching $93.9$\% accuracy, while alternative encodings (\texttt{WAV16} and \texttt{MP3}) show only marginal variations. The dynamic sonification configuration, which incorporates early-execution memory artifacts, reaches $98.0$\% accuracy, outperforming both static configurations and previously published approaches. In comparison, image-based malware detection using RGB feature fusion reports $97.25$\% accuracy, and the BlockDroid CNN model evaluated on a balanced subset achieves $97.38$\%. The results indicate that dynamic memory sonification not only matches but slightly surpasses existing high-performing techniques on the same dataset. Importantly, the improvement is achieved without explicit semantic feature engineering, suggesting that structural patterns embedded in runtime memory provide a highly discriminative signal representation. These findings support the effectiveness of sonification as a competitive alternative to traditional static, image-based, and DL approaches for Android malware detection.

\subsection{Feature Representations and Classification Models}

In addition to comparing memory-region selection strategies and audio encodings, we evaluated different feature extraction paradigms and classification models in order to understand their relative impact on performance. Two complementary feature representations were considered. The first approach relied on manually extracted audio descriptors, including spectral, cepstral, energy-based, and tonal features. These descriptors summarize the signal in a compact tabular form and were used as input to traditional ML classifiers such as Logistic Regression, Support Vector Machines, Random Forest, boosting methods, and ensemble models. This approach provides interpretability and direct control over the extracted signal characteristics. The second approach leveraged deep representations through a one-dimensional Convolutional Neural Network (CNN1D) and transformer-based embeddings. The CNN1D operates directly on structured signal inputs and automatically learns hierarchical filters that capture local temporal and frequency patterns without requiring handcrafted descriptors. In this setting, early convolutional layers capture low-level signal regularities, while deeper layers model more abstract structural relationships within the sonified memory. Across dynamic experiments, several classifiers (including Random Forest, boosting methods, CNN1D, and ensemble approaches) converged to identical peak performance when informative memory regions were selected. This indicates that the discriminative information embedded in the sonified runtime memory is sufficiently structured to be captured both by manually engineered descriptors and by automatically learned convolutional filters.

Notably, once optimal memory-region combinations such as \texttt{apk\_full\_stack} were used, differences between classifiers became marginal. In contrast, when less informative regions were selected, performance degradation was observed across all models. This suggests that memory-region selection plays a more decisive role than the specific classifier architecture. Overall, CNN-based approaches confirm that sonified memory contains stable and learnable structural patterns, while feature-based methods demonstrate that these patterns are also captured through compact spectral and cepstral summaries. The convergence of results across modeling paradigms strengthens the robustness of the proposed dynamic sonification methodology.

\subsection{VirusTotal Dynamic Dataset}

In this set of experiments, we downloaded $500$ malware from VirusTotal, all with obfuscation and anti-analysis techniques. For the benign samples, we downloaded $500$ APK from Androzoo, a popular dataset. Apps span from $2020$ to $2024$ targeting different Android API Levels. The main objective of this study is to prove the efficiency of detecting stealthy Android malware through dynamic memory forensics via audio-signal encoding.

\begin{figure}[t]
    \centering
    \includegraphics[width=\linewidth]{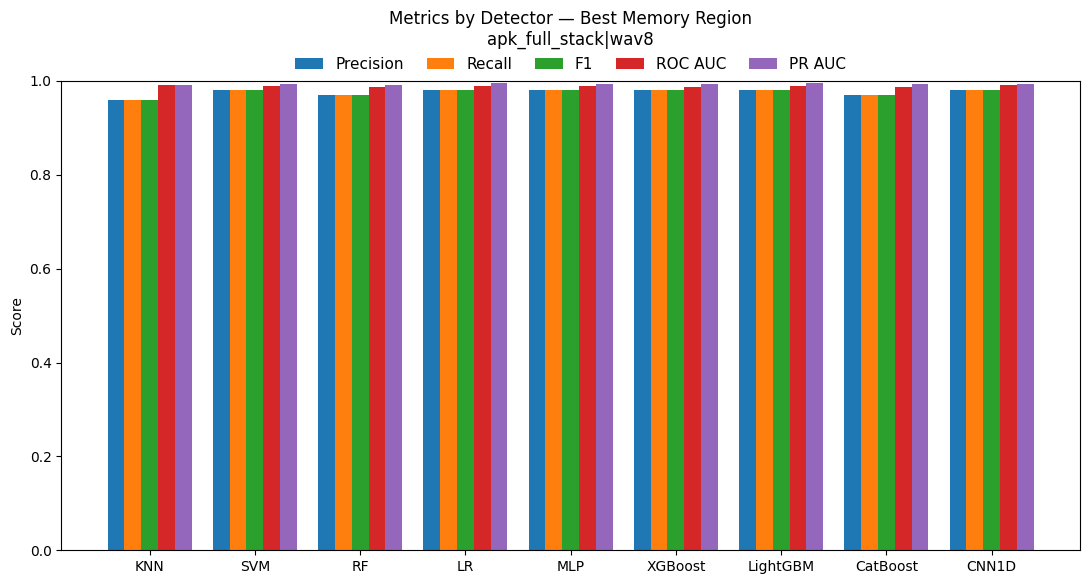}
    \caption{Performances across different detectors in the VirusTotal dataset with the best memory region selection (apk\_full\_stack)}
    \label{fig:detectorsVT}
\end{figure}

The overall comparison among classifiers is reported in Figure~\ref{fig:detectorsVT}. Under the best memory-region configuration (\texttt{apk\_full\_stack} with \texttt{WAV8}), all detectors achieve consistently high performance, with precision, recall, and F1 scores above 0.96 and ROC/PR AUC values approaching 1.0. The highest aggregate performance is obtained by the MLP classifier, which reaches approximately $99.0$\% macro-F1 (about $98.9$\% accuracy). Notably, the performance gap among models is marginal: Random Forest, boosting methods (XGBoost, LightGBM, CatBoost), CNN1D, and SVM converge to nearly identical scores. This limited variance indicates that, once an informative memory-region selection is adopted, classification becomes largely model-independent. In other words, the discriminative strength of the representation dominates over architectural differences among learners.

\begin{figure}[t]
    \centering
    \includegraphics[width=\linewidth]{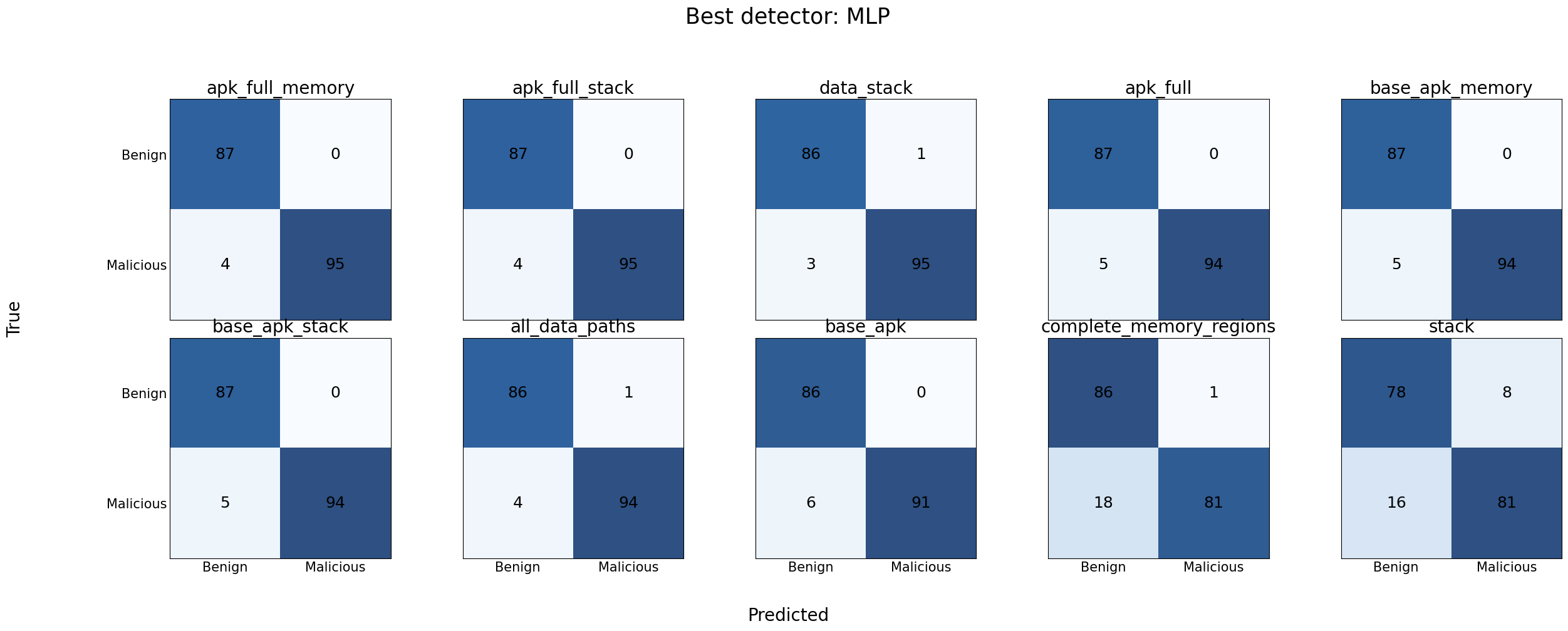}
    \caption{Confusion matrices for the different memory regions selection with the best learning algorithm (MLP)}
    \label{fig:confusionmatrixVT}
\end{figure}

The impact of memory-region selection is illustrated in Figure~\ref{fig:confusionmatrixVT}, where confusion matrices are shown for the best-performing detector (MLP). Several configurations exhibit almost identical confusion matrices, with near-zero (or near-zero) false positives and only four to five false negatives. These strategies share a common structural characteristic: they combine static application code segments (e.g., \texttt{base.apk}, \texttt{.odex}, \texttt{.vdex}) with execution-context information such as stack regions. The stability and peak performance across these configurations suggest that discriminative information emerges from the interaction between static bytecode artifacts and runtime structural patterns. Similarly, \texttt{data\_stack} and \texttt{all\_data\_paths} maintain strong performance, with at most one false positive and three to four false negatives. Although slightly more variable, these results confirm that incorporating runtime-loaded data alongside code segments enhances separability between benign and malicious samples. The consistency across these hybrid strategies indicates that malware-related structural regularities are distributed across multiple execution-aware regions. In contrast, isolated or overly inclusive memory selections lead to a measurable performance degradation. The stack configuration yields a noticeable increase in both false positives and false negatives, indicating that execution context alone does not provide sufficient structural completeness for robust discrimination. Likewise, the \texttt{complete\_memory\_regions} strategy, despite aggregating all mapped segments, introduces substantial noise, as evidenced by the higher number of false negatives. This behavior suggests that indiscriminate inclusion of heap and auxiliary regions dilutes discriminative patterns rather than reinforcing them. The \texttt{base\_apk} configuration alone remains competitive but slightly underperforms compared to its stack-augmented counterpart, reinforcing the importance of integrating runtime context with static code.

Overall, the results shown in Figures~\ref{fig:detectorsVT} and \ref{fig:confusionmatrixVT} indicate that memory-region selection plays a more critical role than the specific learning architecture. When static and dynamic components are jointly considered, the resulting representation becomes sufficiently expressive to allow multiple classifiers to converge to near-identical, near-optimal performance. These findings support the hypothesis that structural regularities distinguishing benign from malicious applications are not confined to a single memory component, but instead arise from the coupling of code artifacts and execution-state information.

Regarding the five misclassified samples, we checked them across all memory regions and classifiers. We focused our attention on two main samples reported in the following.

The first misclassified sample\footnote{SHA256: 051eedb48bed1374a5acc7418d3ac2392450f7fdf548bfd4cc628111283931e5} is detected by 30/67 security vendors and labeled as \textit{trojan}, \textit{PUA}, or \textit{siggen}. Static and behavioral reports highlight obfuscation, reflection usage, runtime modules, telephony access, GPS checks, and even invocation of \texttt{su} commands. At the same time, the application corresponds to a modified version of a commercial VPN app and embeds numerous legitimate SDKs (e.g., Firebase, Facebook SDK, Google Ads, Yandex services). This dual nature, combining legitimate networking functionality with intrusive or repackaged components, places the sample in a gray area between riskware and trojanized software. From a structural perspective, the extensive presence of third-party libraries and standard networking modules likely introduces benign memory patterns that partially mask malicious traits. As a result, the sonified memory representation may resemble that of legitimate, ad-supported applications, explaining the misclassification.

The second misclassified sample\footnote{SHA256: 187fa5b650cb11ce05a9e7c004a8d5583df64e29acee9035d89d7b644a22060e} is consistently labeled as \textit{trojan.spyware} (SmsSpy / SmsEye) by 26/58 vendors. Detection names explicitly reference SMS spying behavior, and the application requests \texttt{RECEIVE\_SMS} permission while registering a receiver for \texttt{android.provider.Telephony.SMS\_RECEIVED}, confirming its spyware intent. Unlike the previous case, however, this APK is relatively compact (6.4 MB) and exhibits a limited number of embedded domains and bundled components. Its implementation appears minimalistic, without heavy packing, native loaders, or complex runtime modules. Consequently, its memory footprint is structurally simpler and may lack the richer entropy patterns typically captured by the sonification-based features. This reduced structural complexity likely decreases separability in the selected memory regions, leading to misclassification despite its clearly malicious functionality.

\subsection{Discussion}

The comparison between static and dynamic sonification highlights the importance of runtime artifacts in shaping the signal representation. Static analysis processes only the packaged bytecode contained in \texttt{classes.dex}. In contrast, dynamic memory snapshots include execution stack frames, dynamically loaded libraries, in-memory resources, and potentially decrypted or unpacked payloads. These elements are not always observable through static inspection and therefore introduce additional structural variability in the binary stream prior to sonification. Spectrogram analysis revealed distinct structural differences between benign and malicious dynamic samples. Malicious applications exhibited persistent horizontal frequency bands concentrated in specific ranges, whereas benign samples showed more uniform spectral distributions across time and frequency. These differences suggest that runtime memory introduces structured and repeated byte sequences that manifest as stable spectral signatures in the time–frequency domain. Regarding encoding formats, \texttt{WAV8} preserves a direct one-to-one mapping between each byte and an 8-bit amplitude sample, maintaining a compact and structurally faithful representation of the binary stream. \texttt{WAV16} increases amplitude resolution by representing samples on 16 bits, effectively expanding dynamic range; however, since the original binary data are inherently 8-bit values, this increase does not introduce additional structural information and may instead smooth or redistribute energy without enhancing discriminability. \texttt{MP3} encoding applies perceptual compression, removing frequency components deemed less relevant for human hearing and introducing quantization artifacts. Despite this transformation, classification performance remains comparable under optimal configurations, indicating that malware-discriminative information is encoded in robust structural patterns rather than in fine-grained waveform precision. Overall, dynamic sonification enhances separability compared to static sonification because runtime memory captures execution-dependent artifacts that introduce additional structured patterns not present in packaged bytecode alone.

\begin{figure}[t]
    \centering
    \begin{subfigure}[b]{0.48\textwidth}
        \includegraphics[width=\textwidth]{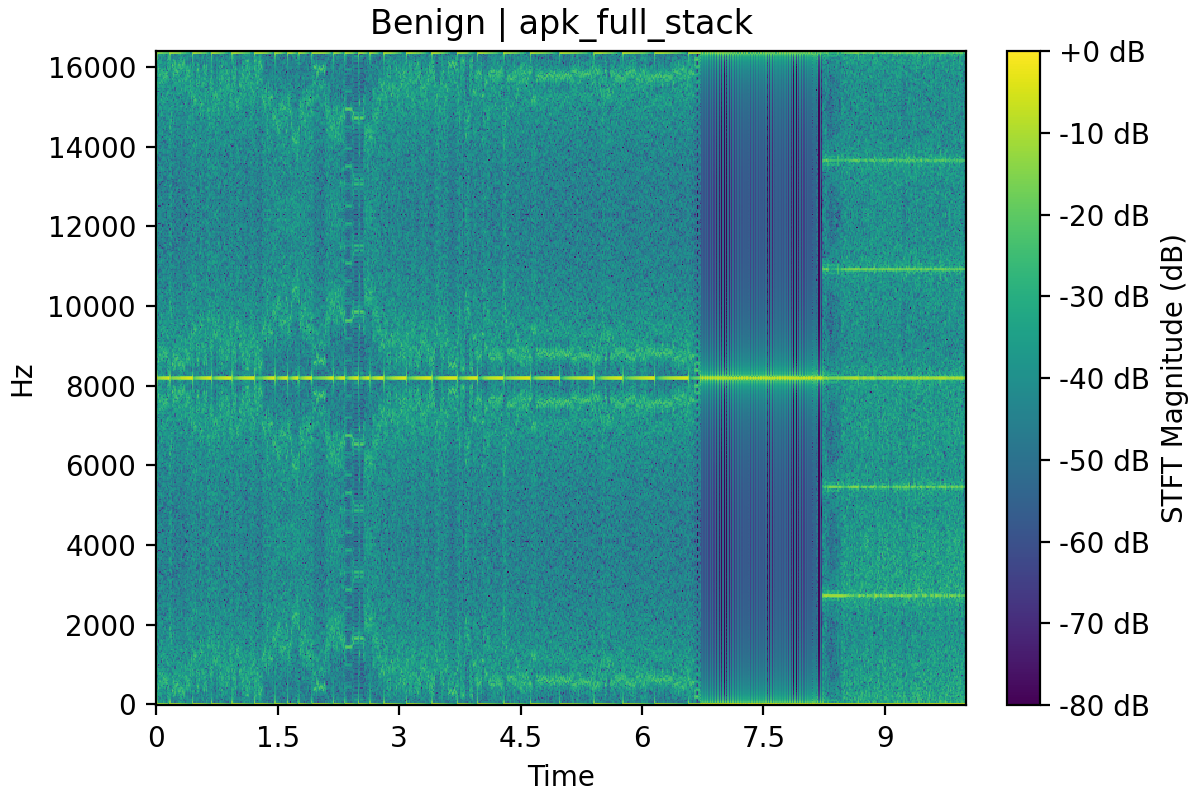}
        \caption{Benign sample}
        \label{fig:spectrogram_benign}
    \end{subfigure}
    \hfill
    \begin{subfigure}[b]{0.48\textwidth}
        \includegraphics[width=\textwidth]{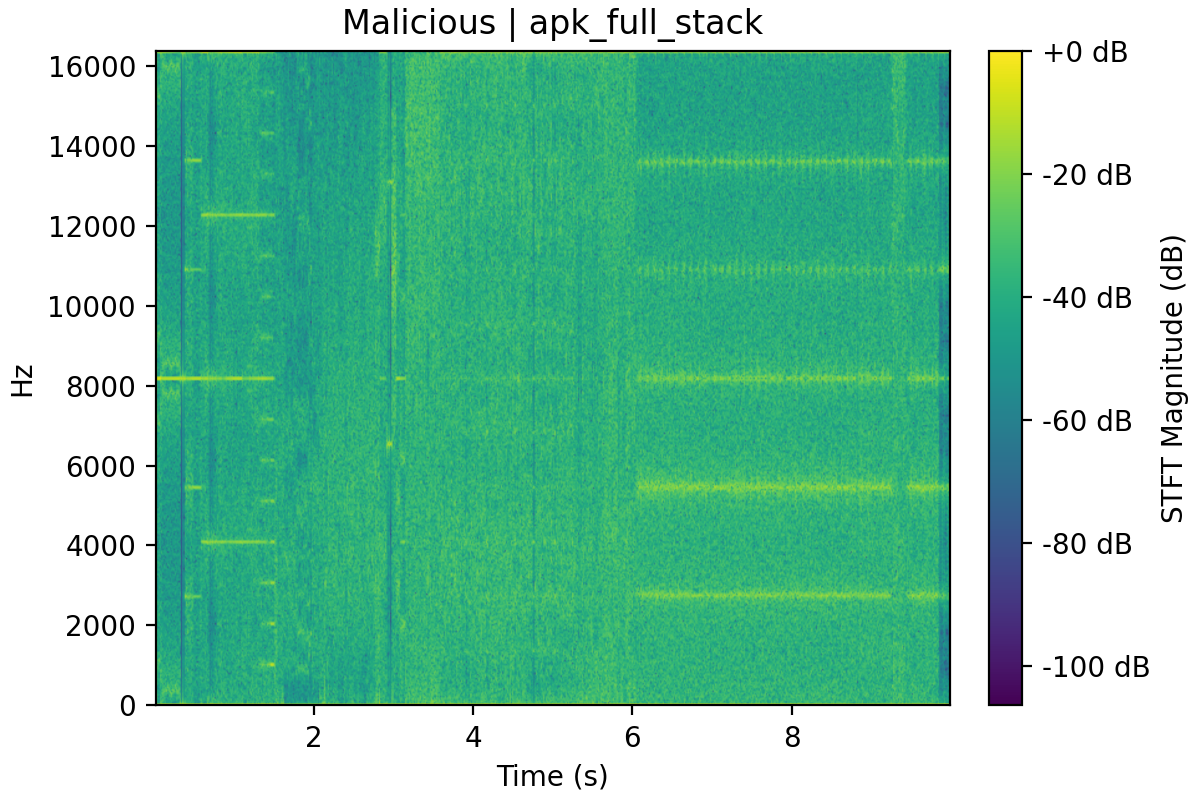}
        \caption{Malicious sample}
        \label{fig:spectrogram_malicious}
    \end{subfigure}
    \caption{STFT spectrogram comparison from dynamic dataset using WAV8 encoding and \texttt{apk\_full\_stack} memory region. Left: benign sample. Right: malicious sample }
    \label{fig:spectrogram_comparison}
\end{figure}

Figure~\ref{fig:spectrogram_comparison} presents STFT spectrograms comparing a benign and a malicious sample using the \texttt{apk\_full\_stack} strategy with \texttt{WAV8} encoding. The malicious sample in Figure~\ref{fig:spectrogram_malicious} exhibits distinct horizontal frequency bands concentrated at approximately $2$--$4$ kHz, $8$ kHz, and $12$--$14$ kHz throughout the temporal domain. In contrast, the benign sample in Figure~\ref{fig:spectrogram_benign} displays a more uniform spectral distribution across frequencies. These structural differences in the frequency domain help explain how machine learning models effectively distinguish malicious from benign applications, achieving $98.0$\% classification accuracy.